\documentclass[aps,prl%
,twocolumn%
,showpacs%
,superscriptaddress%
]{revtex4}

\usepackage{graphicx}
\usepackage{times}

\begin{document}

\setlength{\textfloatsep}{1.2\parindent plus 2pt minus 2pt}
\setlength{\arraycolsep}{1pt}
\newcommand{\be}{\begin{equation}}
\newcommand{\ee}{\end{equation}}
\newcommand{\bea}{\begin{eqnarray}}
\newcommand{\eea}{\end{eqnarray}}
\newcommand{\ac}{{\em ac} }
\newcommand{\ie}{{\it i.e.}}
\newcommand{\eg}{{\it e.g.}}

\newcommand{\affila}{
 Dept. of Physics, University of Notre Dame,
 Notre Dame, IN 46566, USA
}
\newcommand{\affilb}{
 Dept. of Biological Physics, E\"otv\"os University,
 P\'azm\'any P. stny. 1A, H-1117 Budapest, Hungary
}

\title{
 Two-dimensional particle motion in a random potential under \ac bias
}

\author{Maxim A. Makeev}
 \email{makeev@usc.edu}
 \affiliation{\affila}
\author{Imre Der\'enyi}
 \email{derenyi@elte.hu}
 \affiliation{\affilb}
\author{Albert-L\'aszl\'o Barab\'asi}
 \email{alb@nd.edu}
 \affiliation{\affila}

\date[]{\protect\today}

\begin{abstract}
We study the Brownian motion of a single particle coupled to an
external \ac field in a two-dimensional random potential. We find
that for small fields a large-scale vorticity pattern of the
steady-state net currents emerges, a consequence of local symmetry
breaking. In this regime the net currents are highly correlated,
the spatial correlation function follows a logarithmic dependence,
and the correlation length is of the order of the system size. For
large external fields correlations disappear and only random net
currents are observed. Numerical analysis indicates
that the correlation length scales as a power law with both the
size of the system and the amplitude of the \ac field.
\end{abstract}


\maketitle

  Driven diffusion in random media is a much studied
problem impacting on a number of various fields. By now, it is
well established that the interplay between the annealed
randomness of the diffusion process and the quenched randomness of
the media gives raise to unexpected scaling phenomena that have
been extensively studied in recent decades \cite{ale,hav,bou}. The
range of applicability of the problem of driven diffusive motion
in random media covers various fields, with relaxation phenomena
in spin glasses \cite{nat}, dislocation motion in disordered
crystals \cite{hir}, transport in porous media \cite{sta}, and
turbulent diffusion \cite{rich} being but a few typical examples.
In the presence of an external \ac field, the motion of a particle
is the result of the combined effect of thermally activated
diffusivity and periodic motion, induced by the coupling to the
applied field. At time scales much larger than the period of the
driving force, one would expect that the influence of the field is
negligible and the dynamics of the system can be described by the
classical Brownian motion. Indeed, while the particle moves back
and forth along the direction of the \ac field, a stroboscopic
view of the system, obtained by taking pictures only at times that
are integer multiples of the external field period, would still
show a randomly diffusing particle, almost as if the external
field was absent in the system.  In contrast to this intuitive
picture, we demonstrate that the presence of an external ac field
can fundamentally change the nature of the dynamics, when quenched
randomness is present in the system.

An indication,
that an \ac field bias may drastically influence the nature
of diffusive motion in a random potential comes from the
recent advances in the field of thermal ratchets dealing
with one-dimensional (1D) systems, where the $x \rightarrow -x$
symmetry is broken \cite{ratchet}. In equilibrium, a particle
moving in a periodic asymmetric potential displays
a simple diffusive behavior.
However, if the particle is also driven by an \ac
field, it drifts in the direction defined by the
asymmetry of the potential, on time scales exceeding
the period of the \ac field. The appearance of this
non-equilibrium steady-state net current is called the {\em
ratchet effect}, and the average drift velocity of the
particle is often referred to as the {\em ratchet velocity}
\cite{ratchet}. It must be noted, however, that the ratchet
velocity is expected to vanish when the particle is moving
in a quenched random potential, since such a potential obeys
inversion symmetry (in a statistical sense). In particular,
the ratchet velocity in a system of identical energy wells
separated by high enough energy barriers is always zero,
irrespective of the distribution of the barrier heights.

On the other hand, in two-dimensional (2D) systems any {\em 
finite} region of a random potential exhibits some degree of
broken symmetry, and the particle can follow different 
trajectories during the two half periods of the \ac driving.
In this Letter, we show that the interplay between the local
symmetry breaking in a random 2D potential landscape and the
motion of the particle biased by an external \ac field leads
to the appearance of a highly correlated steady-state net 
current field, which is characterized by a large-scale 
vorticity. We also investigate the scaling properties of 
these fields.

\begin{figure}[!b]
\centerline{\includegraphics[width=7.0cm]{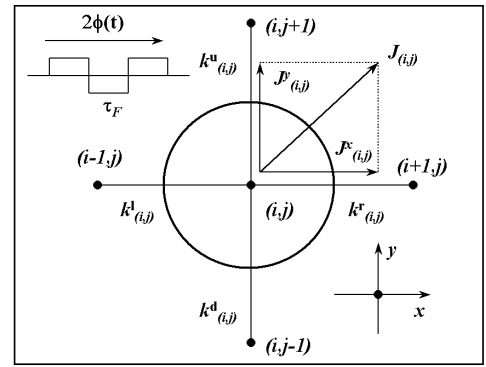}}
\caption{
Schematic illustrating the system under consideration.
A particle on the site $(i,j)$ hops to its neighboring sites,
with hopping rates, $k^{\gamma}_{i,j}$, defined by both the
potential barriers between these sites and by the \ac field.
}
\label{fig1}
\end{figure}

As a model system, we investigate the driven dynamics of a 
single particle moving in a random uncorrelated potential on
a 2D square Euclidean lattices $(i,j)$ with varied number of
identical sites $N = L \times L$, where each site is connected
to its nearest neighbors by bonds of unit length. Random 
potential barriers are assigned to each bond and periodic 
boundary conditions are assumed. A schematic illustration of 
the system's geometry is shown in Fig.\ \ref{fig1}. The 
particle, diffusing in the system, is also driven by an 
external \ac field, applied along the $x$-direction. We define
the local probability currents $J^{x}_{i,j}$ [$J^{y}_{i,j}$] 
on site $(i,j)$ as the currents flowing over the potential 
barriers between the lattice sites $(i,j)$ and $(i+1,j)$ 
[$(i,j+1)$]. As shown in Fig.\ \ref{fig1}, a particle, located
at site $(i,j)$ at an arbitrary moment of time $t$, can hop to
any of the {\em four} nearest neighbor lattice sites by 
overcoming the potential barriers, $E_{i,j}^\gamma$, assigned
to each bond. To simplify the notations, we denote by
$E^{\rm r}_{i,j}$, $E^{\rm l}_{i,j}$, $E^{\rm u}_{i,j}$ and
$E^{\rm d}_{i,j}$ the local potential barriers associated with
the right
$E^{\rm r}_{i,j} = E_{(i,j) \rightarrow (i+1,j)}$, left
$E^{\rm l}_{i,j} = E_{(i,j) \rightarrow (i-1,j)}$, up
$E^{\rm u}_{i,j} = E_{(i,j) \rightarrow (i,j+1)}$, and down
$E^{\rm d}_{i,j} = E_{(i,j) \rightarrow (i,j-1)}$ jumps, such
that $E^{\rm r}_{i,j} \equiv E^{\rm l}_{i+1,j}$ and
$E^{\rm u}_{i,j} \equiv E^{\rm d}_{i,j+1}$.
The barrier heights are chosen to be random
and quenched, with the spatial distribution given by
$E^{\gamma}_{i,j} = E_0 + U \eta^{\gamma}_{i,j}$, where 
$\gamma$ stands for r, l, u, and d; $E_0$ is a constant; and
$\eta^{\rm r}_{i,j} \equiv \eta^{\rm l}_{i+1,j}$ and
$\eta^{\rm u}_{i,j} \equiv \eta^{\rm d}_{i,j+1}$
are uncorrelated random numbers
uniformly distributed between 0 and 1. Throughout the paper we
measure the energies in units of $k_{\rm B}T$, where $k_{\rm 
B}$ stands for the Boltzmann constant, and $T$ is the absolute
temperature.

In addition to the thermally activated diffusion, the particle
is also driven by an external \ac force, $2\phi(t)$, applied along
the $x$-direction (see Fig.\ \ref{fig1}). Since the energy barriers
are placed halfway between adjacent sites, the \ac force modulates
their hight by $-\phi(t)$ in the positive $x$-direction and by 
$+\phi(t)$ in the negative one. Thus, for large enough barriers, 
the hopping rates of the particle occupying site $(i,j)$ can be 
written as
\bea
k^{\rm r}_{i,j} & = &
\nu \exp \{ - [ U \eta^{\rm r}_{i,j} - \phi(t) ] \},
\nonumber\\
k^{\rm l}_{i,j} & = &
\nu \exp \{ - [ U \eta^{\rm l}_{i,j} + \phi(t) ] \},
\nonumber\\
k^{\rm u}_{i,j} & = &
\nu \exp \{ -   U \eta^{\rm u}_{i,j} \},
\nonumber\\
k^{\rm d}_{i,j} & = &
\nu \exp \{ -   U \eta^{\rm d}_{i,j} \},
\label{eq1}
\eea
where $\nu = \nu_0 \exp \{ -E_0 \}$, and the attempt frequency
$\nu_0$ is also a constant.

Next, we turn to an ensemble description, in which the probability
of the particle occupying site $(i,j)$ is denoted by $P_{i,j}$, 
and the time evolution of this probability distribution is 
described by the following set of master equations:
\be
\frac{\partial P_{i,j}}{\partial t}  =
-J^x_{i,j}+J^x_{i-1,j}-J^y_{i,j}+J^y_{i,j-1},
\label{eq2}
\ee
where
\bea
J^x_{i,j} & = & k^r_{i,j}P_{i,j} - k^l_{i+1,j}P_{i+1,j},
\nonumber\\
J^y_{i,j} & = & k^u_{i,j}P_{i,j} - k^d_{i,j+1}P_{i,j+1}
\label{eq3}
\eea
are the $x$ and $y$ components of the local probability currents
${\vec J}_{i,j}$, as defined above.

For simplicity, in the following we restrict ourselves to the
consideration of symmetric square wave external fields, i.e, when
the field $\phi(t)$ alternates between $+F$ and $-F$ at constant
time intervals $\tau_{F}$. We also assume that $\tau_{F}$ is much
larger than the relaxation time of the entire system, estimated 
as $\tau_{\rm relax} \approx L^2 \max(1/k^\gamma_{i,j})$.
In this case, for each half period of the \ac driving the 
probability distribution $P_{i,j}$ and currents ${\vec J}_{i,j}$ 
relax to their steady state values: $P_{i,j}(+F)$, 
${\vec J}_{i,j}(+F)$, and $P_{i,j}(-F)$, ${\vec J}_{i,j}(-F)$), 
which can be determined from the stationary solution of the master
equation for $\phi(t)=+F$ and $\phi(t)=-F$, respectively. From 
direct analogy with the ratchet velocities, we can then define the
{\em net currents} in the system as
\be
{\vec{\cal J}}_{i,j} =
 \frac{1}{2} [ {\vec J}_{i,j}(+F) + {\vec J}_{i,j}(-F) ].
\label{eq4}
\ee
The appearance of non-zero net currents is a characteristics 
of systems with locally broken symmetry, and their magnitudes
are given by higher than linear terms in the response
functions (\ie, by $F^2$ and higher order terms)
\cite{mdbu}.

\begin{figure}[!t]
\centerline{\includegraphics[width=6.5cm]{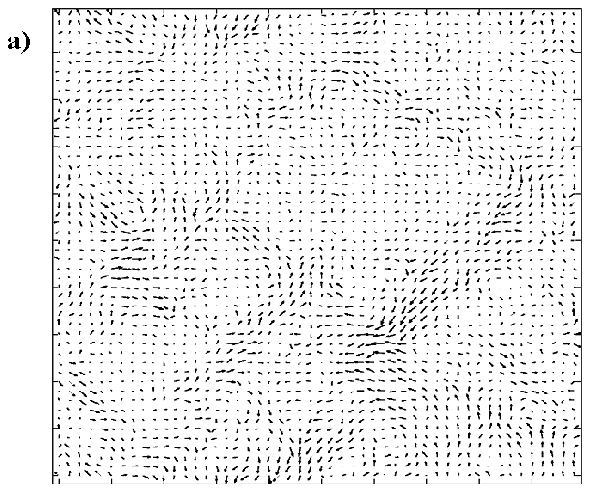}}
\centerline{\includegraphics[width=6.5cm]{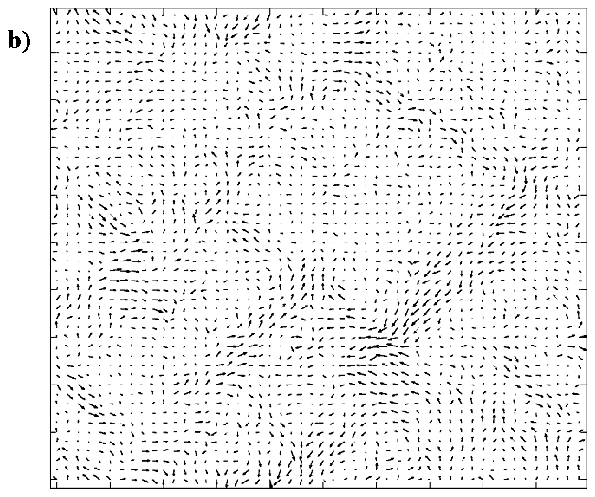}}
\centerline{\includegraphics[width=6.5cm]{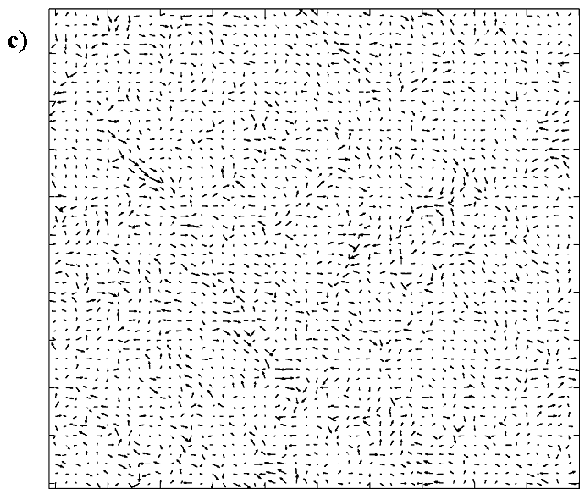}}
\centerline{\includegraphics[width=6.5cm]{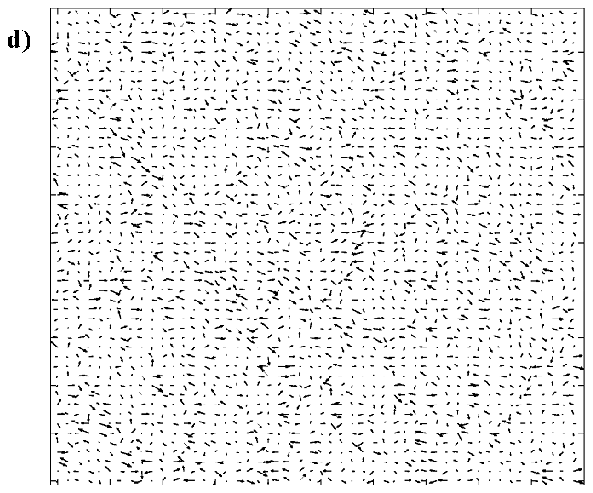}}
\caption{
Snapshots of the steady-state net current
fields obtained for model systems with $L =
50$, $U = 0.5$, and different external field
amplitudes: (a) F = 0.01; (b) F = 0.05; (c) F =
0.10; (d) F = 0.90.
}
\label{fig2}
\end{figure}

We solved the system of Eqs.\ (\ref{eq1})-(\ref{eq3}) numerically
by using the conjugate gradient method, modified for sparse
matrices \cite{teuko}. In Figs.\ \ref{fig2} (a)-(d), we show the
steady state local net current patterns obtained for a system with
linear size $L = 50$, randomness parameter $U = 0.5$, and external
field amplitudes $F = 0.01$, 0.05, 0.1, and 0.9. These figures
offer a visual proof for the emergence of highly non-trivial
steady-state net current fields, characterized by long-range
correlations and large-scale vorticity. The vorticity structures
are most apparent for small external field amplitudes ($F \ll U$).
As $F$ increases, however, the correlations gradually disappear
and the net current field converges to a nearly random structure.

It must be emphasized that numerical solution of Eqs.\
(\ref{eq1})-(\ref{eq3}) for arbitrarily small values of $F$ and/or
for arbitrarily large system sizes is not possible. This is due to
the slow convergence of the conjugate gradient method in these
limits and, most importantly, due to the growing impact of the
randomization, induced by the numerical errors \cite{mdbu}.

   To understand the nature of the observed correlations
and investigate the scaling properties of the system, we
compute the ensemble-averaged current-current correlation
function, defined in the real space as
\be
C(r)=
 \left<
  \frac{1}{\sum_{(i,j)} {\vec{\cal J}}_{i,j}^2 N_{r}}
  \sum_{i,j} \sum_{i',j'}
  \left(
   {\vec{\cal J}}({\bf r}_{i,j}) - {\vec{\cal J}}({\bf r}_{i',j'})
  \right)^2
 \right>.
\label{eq5}
\ee
Here summation over all the lattice site indexes
$(i,j)$ is implied, while  summation over $(i',j')$ indexes is
performed only for lattice site pairs such that $|{\bf
r}_{i,j}-{\bf r}_{i',j'}| = r$, $N_{r}$ being the number of such
pairs. The averaging $<...>$ was taken for various different
realizations of the disorder $\{\eta_{i,j}^\gamma\}$. The
correlation function defined this way is bounded from below by 0
(in case of perfect correlation), from above by 4 (perfect
anti-correlation), and takes the value of 2 for vanishing
correlation.

\begin{figure}[!b]
\centerline{\includegraphics[width=7.0cm]{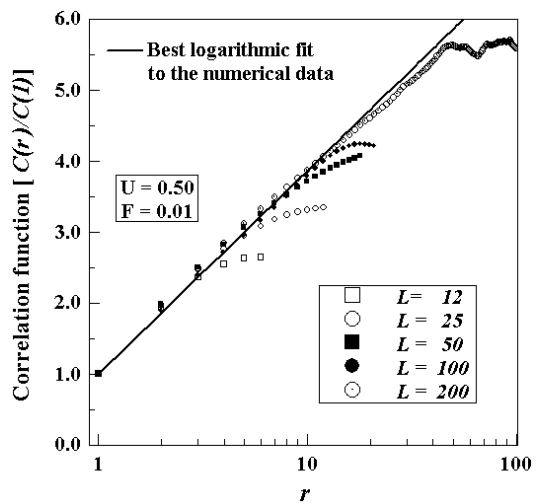}}
\caption{
Normal-log plot of the current-current correlation
function $C(r)/C(1)$, computed for $U = 0.5$ and $F =
0.01$ for different system sizes.
}
\label{fig3}
\end{figure}

In Fig.\ \ref{fig3}, we show the normalized correlation function,
$C(r)/C(1)$, computed for different system sizes, and fixed
external field amplitude, $F$ = 0.01. As Fig.\ \ref{fig3}
demonstrates, the correlation function follows closely a
logarithmic dependence on $r$, for small radial distances, and
deviates from that as $r$ grows beyond a certain value, with a
saturation threshold determined by the system size, $L$. This
allows us to introduce a correlation length $\xi$ in the system,
which we define as a characteristic length at which the
correlation vanishes: $C(r)=2$, that is, $1 - 0.5\;C(r) \sim
<{\vec{\cal J}}({\bf r}_{i,j}) {\vec{\cal J}}({\bf r}_{i',j'})>$
turns zero.

\begin{figure}[!t]
\centerline{\includegraphics[width=7.0cm]{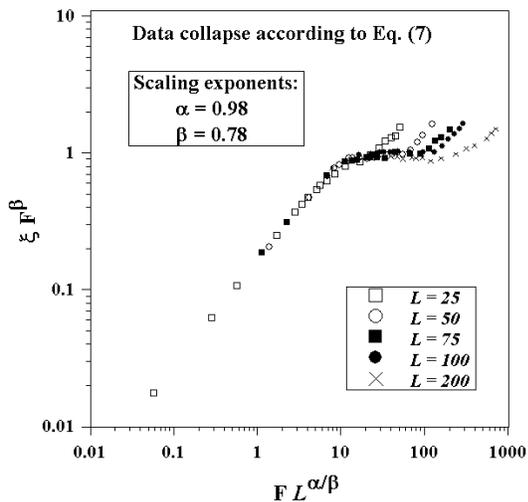}}
\caption{
Scaling plot of the correlation length according
to Eq.\ (\ref{eq7}) for different system sizes.
}
\label{fig4}
\end{figure}

 On the basis of the above observations, we conclude that,
in the most general form, the current-current correlation function
is described well by the relation
\be
C(r) \sim \log(r) f\left(\frac{r}{\xi}\right),
\label{eq6}
\ee
where $f(x \le 1) \sim$ const.
Note that the correlation length in Eq.\ (\ref{eq6})
is dependent on both the amplitude of the external field and the
system size; \ie, $\xi = \xi(F,L)$. Moreover, as our results
demonstrate, the correlation length, $\xi$, scales with both of
these quantities following power laws. Thus, for small amplitudes
of the external field, the correlation length depends strongly on
the system size, approximately following $\xi(L) \sim L^{\alpha}$
relation, where the exponent $\alpha = 0.983 \pm 0.008$ (\ie,
$\xi$ grows nearly linearly with $L$). On the other hand, for
intermediate external field amplitudes, $\xi$ is found to be
weakly dependent on $L$ and to monotonically decrease with $F$,
following closely the inverse power law, $\xi \sim F^{-\beta}$,
where the exponent is $\beta \simeq 0.78 \pm 0.01$. The 
transition between the two regimes occurs at a certain field 
amplitude, $F^{\rm c}_1(L)$, dependent on $L$. The above 
scaling laws can be collapsed into a single scaling relation:
\be
\xi \sim F^{-\beta} g(F L^{\alpha/\beta}),
\label{eq7}
\ee
where $g(x) \sim x$ for $F \le F^{\rm c}_1(L)$ and
$g(x) \sim$ const for $F^{\rm c}_1(L) \le F \le F^{\rm c}_2(L)$.
The significance of $F^{\rm c}_2(L)$ will be discussed later.
In Fig.\ \ref{fig4} we show the
numerical data collapse, performed according to Eq.\ (\ref{eq7}).
As one can observe, it provides satisfactory results for small and
intermediate $F L^{\alpha/\beta}$. In general, three regimes of
the scaling behavior can be discriminated. For small $F$, the
behavior is defined by the system size scaling and extends over
the region of $F$ values up to $F^{\rm c}_1(L)$. In the second
regime, the scaling is dominated by the external field amplitude.
This regime corresponds to the intermediate values of $F^{\rm
c}_1(L) \le F \le F^{\rm c}_2(L)$. Note that substantial
deviations from the scaling behavior predicted by Eq.\ (\ref{eq7})
are observed for large values of $F$ (\ie, for $F \ge F^{\rm
c}_2(L))$. This behavior is an artifact of the discrete nature of
the model system under consideration: since the particle is on the
lattice, we cannot measure $\xi$ smaller than 1. As a result, in
the limit of large $F$, the correlation length does not follow
Eq.\ (\ref{eq7}) down to arbitrary small distances, but rather
saturates at $\xi \simeq 1$.

  In summary, we have studied the Brownian motion of a single
particle on 2D Euclidean lattices of different sizes with
quenched random potential, coupled to an
external \ac field. We have found that the interplay between
the quenched randomness of the potential landscape and the
external \ac field bias leads to the emergence of large-scale
vorticity patterns in the net steady-state currents.
These currents are found to be strongly correlated, with the
two-point correlation function following a logarithmic dependence.
The properties of the current fields are
observed to be largely independent of the particular realization
of the disorder. The correlation length has been found to scale
nearly linearly with the system size and display an inverse power
law behavior with the external field amplitude.

An intuitive explanation for the appearance of system-wide structures
is that in 2D the particle has the opportunity to ``choose'' globally
different paths when driven in the two opposing directions. The
superposition of these globally different flow fields results in
large-scale flow patterns. This picture is also consistent with a slow
(logarithmic) decay of the correlations and a correlation length being
in the order of the system size. It is important to note that the above
phenomenon cannot occur in 1D, where all the energy barriers are in
series, and no alternative trajectories are possible.

\begin{acknowledgments}
Research at Notre Dame was supported by NSF-PHYS.
\end{acknowledgments}

\end{document}